\documentclass[usenatbib]{mn2e}
\usepackage[pdftex]{graphicx}
\usepackage{amssymb}
\usepackage{amsmath}
\usepackage{verbatim}
\usepackage{array}

\def\gtsima{$\; \buildrel > \over \sim \;$}
\def\gtsim{\lower.5ex\hbox{\gtsima}}

\def\beq{ \begin{equation} }
\def\eeq{ \end{equation} }

\begin{document}


\title{A Self-Consistent Reduced Model for Dusty Magnetorotationally Unstable Discs}

\author[Jacquet \& Balbus]{Emmanuel Jacquet,$^1$ Steven Balbus$^{2,3}$\\
$^1$Laboratoire de Min\'{e}ralogie et de Cosmochimie du Mus\'{e}um, Mus\'{e}um National d'Histoire Naturelle, 57 rue Cuvier, 75005 Paris, France\\
$^2$Laboratoire de Radioastronomie, \'Ecole Normale
Sup\'erieure, 24 rue Lhomond, 75231 Paris CEDEX 05, France\\
$^3$Institut universitaire de France,
Maison des Universit\'es, 103 blvd.\ Saint-Michel, 75005
Paris, France
}





\maketitle

\begin{abstract}
The interaction between settling of dust grains and magnetorotational instability (MRI) turbulence in protoplanetary disks is analyzed. We use a reduced system of coupled ordinary differential equations to represent the interaction between the diffusion of grains and the inhibition of the MRI. The coupled equations are styled on a Landau equation for the turbulence and a Fokker-Planck equation for the diffusion. The turbulence-grain interaction is probably most relevant near the outer edge of the disk's quiescent, or ``dead'' zone. Settling is most pronounced near the midplane, where a high dust concentration can self-consistently suppress the MRI. Under certain conditions, however, grains can reach high altitudes, a result of some observational interest. Finally, we show that the equilibrium solutions are linearly stable. 
\end{abstract}



\section{Introduction}

  The presence of small dust grains in protostellar disks is
critical to the thermal, dynamical, and chemical behavior of the gas, and is a crucial observational diagnostic.  One particularly important
feature is that dust tends to
stabilize disks against the magnetorotational instability (MRI),
and to otherwise complicate our understanding of MHD processes
in such systems \citep{Stoneetal2000,Sanoetal2000,SalmeronWardle2008,BaiGoodman2009}.  By readily adsorbing free electrons onto their surfaces the grains become charge carriers of very low mobility.
In addition, the grains deplete the gas of alkali metal atoms, which
are ordinarily a low ionization potential source of electrons.
This causes the resistivity of the gas-dust mixture to rise dramatically and essentially suppress MRI-powered turbulence over a wide range of heliocentric distances. Indeed, inclusion of dust grains in resistivity calculations result in greater ``dead zones'' than predicted by gas-phase chemical networks alone \citep{Sanoetal2000,BaiGoodman2009}. 

  Most studies thus far have assumed that dust grains were well-mixed over the vertical thickness of the disk \citep{Sanoetal2000,SalmeronWardle2008,BaiGoodman2009}. This is justified for small grains, which are tightly coupled to the gas. However, as grains grow in size, while still controlling the ionization of the gas, and gas becomes less dense, significant decoupling should occur, particularly at large heliocentric distances. In particular, if the gas is stabilized,
the embedded grains will settle toward the disk midplane.
At this stage, depleted of its grains, the ionization rises and
the gas may once again be
vulnerable to the MRI. However, the ensuing turbulent agitation would 
{\em restir} the grains, diffusing them once again upward into the gas
\citep[e.g.][]{Carballidoetal2006}. Of course,
the grains would then suppress the same instability that allowed them
to diffuse through the gaseous envelope.  The ``cycle of inconsistency''
would continue...

How is this behavior ultimately resolved?  It is possible to
 envision a middle ground.  In a turbulent medium, the growth of fluctuations
is generally set 
 by a balance between linear (magnetic tension
forces and resistive dissipation) and nonlinear (cascade) processes.
If saturation occurs at 
low amplitudes where nonlinearity is of secondary importance,
the balance may be more simply regulated by a marginal,
near zero, linear growth rate.  The presence
of {\em some} level of turbulent fluctuations would diffuse dust,
and rather than cut-off the instability, such diffusion might regulate
its growth.  Specifically, the turbulence could stir just enough dust
into the gas to ensure marginal growth: an increase in fluctuations
raises the effective resistivity (more dust), a decrease in fluctuations
increases the effective conductivity (less dust).  The question of the
existence and stability of such a dynamical equilibrium is the focus of
the current paper.

Our chosen method of investigation is to construct and study a
reduced system of ordinary
differential equations, designed to reproduce certain key features
of real disks.  Our mathematical problem consists of two
coupled, nonlinear equations, both of which are a common staple
of reduced systems.   The first is a simple nonlinear
Landau equation \citep{LandauLifshitz1959} for the fluctuation
amplitude.  The second is a Fokker-Planck equation,
with both drift and diffusive terms,
for the concentration of dust grains.  The diffusion coefficient
of the Fokker-Planck equation is a function of the fluctuation level,
whereas the growth rate of the Landau equation depends on the
grain concentration.  It is this particular
mathematical coupling in our proxy system that makes it interesting
for astrophysical applications.  We are able to demonstrate
the existence of stable solutions for our reduced systems.  

This article is organized as follows: In Section 2, we investigate the conditions under which the interaction studied is relevant in the disk. In Section 3, we outline our reduced system and its equilibrium solution. Proof of its stability is deferred to Appendix \ref{stability}. In Section 4, we adopt a specific form for the growth rate for illustrative purposes. In Section 5, we conclude.

\section{Dust and MRI: a review}


In this section, we review the role of dust on the MRI in a schematic
way to orient the reader with respect to order-of-magnitude scalings, and
to highlight the conditions under
which the interaction of dust settling with the MRI may be relevant. 
There are three primary criteria for this:
\begin{itemize}
\item[(i)] The dust dominates recombination of ions and electrons.
\item[(ii)] Nonideal MHD effects are important, but do not suppress MHD turbulence altogether.
\item[(iii)] A substantial fraction of the dust grains can settle to the midplane.
\end{itemize}
We shall quantify each of these in the next subsections.

The disk is described in a cylindrical coordinate system, with $R$ the heliocentric distance and
$z$ the height above the midplane.
Since our calculations are local,
we need not specify a global disk model but 
we shall normalize
our results to values of order those of the minimum mass solar nebula \citep[MMSN;][]{Hayashi1981} near an
heliocentric distance of 10 AU. 
We will assume the disk to be axisymmetric and vertically isothermal, with the gas density given by:
\begin{equation}
\rho=\frac{\Sigma}{\sqrt{2\pi}H}\exp{(-\frac{z^2}{2H^2})},
\end{equation}
with $\Sigma$ the surface density, $c_s$
the isothermal sound speed, $H=c_s/\Omega$ is the pressure scale height and $\Omega$ the Keplerian angular velocity. $P=\rho c_s^2$ is the corresponding pressure.                         

\subsection{The ionization fraction}

Consider a gas composed of neutrals, ions and electrons, and a
population of dust grains, of respective number densities $n_n$, $n_i$, $n_e$
and $n_p$, all assumed to be at the same temperature $T$. 
The neutrals are predominantly H$_2$
molecules and the ions are treated as one singly-charged
species. The grains are assumed to be identical spheres
of radius $a$ and internal density $\rho_s$ ; we also denote by $\rho_p=4\pi \rho_s a^3n_p/3$ the dust mass density and by $\epsilon\equiv\rho_p/\rho$ the dust-to-gas mass ratio\footnote{The grains are considered to be sufficiently large to ignore the effects of their electrical charge on their own dynamics and their direct contribution to the current density (see \citealt{Wardle2007}) and thence nonideal MHD terms (but see \citet{Bai2011} for the effects PAH-sized grains).}. We denote by $\zeta$ the ionization rate, which in the outer solar system shall be dominated by cosmic rays \citep{BaiGoodman2009}.  
The evolution equations for $n_e$ and $n_i$ are:
\begin{equation}
\frac{\partial n_e}{\partial t}=\zeta n_n - \beta_{\rm eff}n_en_i - I_e\pi a^2 v_{Te}n_pn_e
\label{dne/dt}
\end{equation}
\begin{equation}
\frac{\partial n_i}{\partial t}=\zeta n_n - \beta_{\rm eff}n_en_i - I_i\pi a^2 v_{Ti}n_pn_i,
\label{dni/dt}
\end{equation}
where $\beta_{\rm eff}$ is the effective gas-phase electron-ion recombination rate,
$v_{Te,i}\equiv \sqrt{8k_BT/\pi m_{i,e}}$ 
is a characteristic thermal speed, 
and $I_{e,i}$ the (averaged) product of the sticking coefficient and the focusing factor due to electrostatic effects (the $\tilde{J}$ of \citealt{DraineSutin1987}).

  In the absence of dust grains, using charge neutrality, the fractional abundance of electrons at equilibrium ionization is \citep{Gammie1996,Fromangetal2002}:
\begin{eqnarray}
x_e\equiv\frac{n_e}{n_n}&=&\sqrt{\frac{\zeta}{\beta_{\rm eff}n_n}}\nonumber\\&=& 2\times 10^{-10}\left(\frac{\zeta}{10^{-17}\:\mathrm{s^{-1}}}\frac{10^{-16}\:\mathrm{m^3/s}}{\beta_{\rm eff}}\frac{10^{-9}\:\mathrm{kg/m^3}}{\rho}\right)^{1/2}
\end{eqnarray}
Since metal ions have a low recombination rate (see Appendix \ref{molecule metal}),  a small fraction of their cosmic abundance is then sufficient for 
$x_e$ to allow widespread MRI activity \citep{Fromangetal2002,BaiGoodman2009} \textit{if dust is ignored}. The appreciable depletion ($\gtrsim$10 \%) of chondrites in moderately volatile elements (e.g. alkalis) relative to the total condensable matter \citep[e.g.][]{ScottKrot2003} suggests that these elements were not very efficiently removed from the \textit{gas} phase. Their depletion in the gas phase likely did not exceed 1-2 orders of magnitude then (similarly to cold interstellar gas, e.g. \citealt{Yin2005}), in contrast to the much stronger depletions envisioned in some parameter studies \citep[e.g.][]{Fromangetal2002,IlgnerNelson2008,Flaigetal2011}. 
Metals \textit{per se} would thus be sufficiently abundant to significantly reduce the extent of the dead zone  \citep{Fromangetal2002,BaiGoodman2009} but this does not hold if dust is taken into account \citep{Sanoetal2000,BaiGoodman2009}. Dust thus appears as the main agent acting to suppress the MRI.

If gas-phase recombination can be neglected (which we seek here to quantify), we have, at equilibrium:
\begin{align}
x_e = \frac{4}{3}\frac{\zeta\rho_sa}{I_e\rho_pv_{Te}} = 3\times 10^{-13}\frac{1}{I_e}\left(\frac{\zeta}{10^{-17}\: \mathrm{s}^{-1}}\right)\left(\frac{\rho_sa}{10^{-2}\:\mathrm{kg/m^2}}\right)\nonumber\\\left(\frac{10^{-11}\:\mathrm{kg/m^3}}{\rho_p}\right)\left(\frac{100\:\mathrm{K}}{T}\right)^{1/2}.
\label{xe}
\end{align}
The normalizing value of $\rho_sa=10^{-2}\:\mathrm{kg/m^2}$ 
applies to micron-sized grains. 
The smaller the grains, the larger the area offered for recombination per unit mass, and hence the lower the ionization fraction. Since the equilibrium attainment timescale  $x_e/\zeta$ is shorter than all other timescales of interest, chemical equilibrium will be assumed throughout. 
 
 Neglect of electron-ion recombination is warranted 
 if: 
\begin{eqnarray}
\frac{\beta_{\rm eff} n_in_e}{\zeta n_n}&=& \frac{16}{9}\frac{\left(m_em_i\right)^{1/2}}{m_{H_2}^2I_eI_i}\frac{\beta_{\rm eff}\zeta}{P}\left(\frac{\rho_sa}{\epsilon}\right)^2\nonumber\\
& \approx & \frac{10^{-4}}{I_eI_i}\left(\frac{\zeta}{10^{-17}\:\mathrm{s^{-1}}}\right)\left(\frac{\beta_{\rm eff}}{10^{-16}\:\mathrm{m^3/s}}\right)\nonumber\\ & & \left(\frac{\rho_sa}{10^{-2}\:\mathrm{kg/m^2}}\frac{10^{-2}}{\epsilon}\right)^2\left(\frac{10^{-3}\:\mathrm{Pa}}{P}\right)\nonumber\\ 
  &\ll &  1.
 \label{gas recombination}
\end{eqnarray}
(Some justification of the normalizing value for $\beta_{\rm eff}$ is provided in Appendix A, where the molecular ion/metal ion ratio is estimated). Under this condition, the dust may be said to control the ionization fraction as required by criterion (i). Note that our focus on dust properties is strictly justified if its impact on the ionization fraction is not overshadowed by vertical variations of the ionization rate, which is satisfied if e.g. the gas column density is smaller than the stopping grammage of the ionizing radiations ($9.6\times 10^2\:\mathrm{kg/m^2}$ for cosmic rays according to \citealt{UmebayashiNakano1981}).

\subsection{MRI activity}
The reduced ionization due to dust enhances nonideal terms in the induction equation, whose importance we now quantify. Various dimensionless numbers have been defined in the literature depending on the diffusivity regime, and the thresholds for MRI activation are still being debated, so we shall restrict ourselves to two of them, pertaining to ohmic and ambipolar diffusion, respectively:

  Ohmic diffusion is believed to dominate near the midplane \citep[e.g.][]{SalmeronWardle2008}. The importance of Ohmic diffusivity $\eta_0$ is measured by the magnetic Reynolds number, which, if we inject equation (\ref{xe}), is given by\footnote{We use $\eta_0=4v_{Te}m_e\sigma_0/(3\mu_0e^2x_e)$ from equations (9), (24) and (25) of \citet{Balbus2009}.
}
\begin{eqnarray}
 \textrm{Re}_M & \equiv & \frac{c_s^2}{\eta_O\Omega}
=\frac{\pi}{8}\frac{\mu_0e^2\sigma_0}{m_{H_2}I_e}\frac{\zeta}{\Omega}\frac{\rho_sa}{\rho_p}\nonumber\\
 &=& \frac{50}{I_e}\left(\frac{R}{10\:\mathrm{AU}}\right)^{3/2}\left(\frac{\zeta}{10^{-17}\:\mathrm{s^{-1}}}\right)\left(\frac{\rho_sa}{10^{-2}\:\mathrm{kg/m^2}}\right)\nonumber\\& & \left(\frac{10^{-11}\:\mathrm{kg/m^3}}{\rho_p}\right).
  \label{ReM}
\end{eqnarray}
with $\sigma_0=10^{-19}\:\mathrm{m^2}$ the neutral-electron 
cross-section \citep{Draineetal1983}. Currently estimated thresholds for $\textrm{Re}_M$ for good coupling between the gas and magnetic fields are $10^{2\pm 2}$ \citep{Fromangetal2002}. 

  The importance of ambipolar diffusion, which may dominate in the upper layers of the disk 
\citep{PerezBeckerChiang2011}, 
is measured by the dimensionless ion-neutral collision rate (per
neutral molecule):
 \begin{eqnarray}
 \textrm{Am} &\equiv& \frac{x_in_n\beta_{\rm in}}{\Omega}
=\frac{4}{3}
 \frac{\beta_{\rm in}\zeta}{I_im_{H_2}v_{Ti}\Omega}\frac{\rho_sa}{\epsilon}\nonumber\\
 &=& \frac{0.5}{I_i}\sqrt{\frac{m_i}{m_{H_2}}}
 \left(\frac{\zeta}{10^{-17}\:\mathrm{s^{-1}}}\right)\left(\frac{\rho_sa}{10^{-2}\:\mathrm{kg/m^2}}\right)\nonumber\\& &\left(\frac{10^{-2}}{\epsilon}\right)\left(\frac{100\:\mathrm{K}}{T}\right)^{1/2}\left(
\frac{R}{10\:\mathrm{AU}}\right)^{3/2},
 \label{Am}
 \end{eqnarray}
with $\beta_{\rm in}=1.9\times 10^{-15}\:\mathrm{m^3/s}$ the ion-neutral collision rate coefficient \citep{Draineetal1983}. 
The threshold for good coupling between ions and neutrals is of order $1-10^2$ \citep{PerezBeckerChiang2011,BaiStone2011}.

  As mentioned previously, other dimensionless numbers exist in the literature 
 but as these are proportional to the ionization fraction, 
their dependence on dust properties is subsumed in the 
factor $\rho_sa/\rho_p$, or equivalently $\rho_sa/\epsilon$. 

  The condition (ii) that the turbulence level is a strong function of ionization is met when the magnetic coupling is marginally good, i.e. when the dimensionless number pertaining to the relevant diffusivity regime is close to its threshold value for ideal MHD. In other words, the region most relevant to our analysis is near the outer edge of the dead zone. From equations (\ref{ReM}) and (\ref{Am}), this edge may be expected to be at $\sim$10 AU from the Sun in a MMSN, as it was in the detailed calculations of \citet{BaiGoodman2009}.


 \subsection{Grain dynamics}

The dynamics of solid grains are dictated by gas drag. For grains small compared to the gas mean free path, the stopping time is \citep{Epstein1924}
 \begin{equation}
 \tau = \sqrt{\frac{\pi}{8}}\frac{\rho_sa}{\rho c_s}.
 \end{equation}
For small grains ($\tau \ll \Omega^{-1}$), the vertical drift velocity of the grains is given by \citep{Dubrulleetal1995}:
\begin{equation}
v(z)=-\Omega^2 \tau z. 
\end{equation}
Settling is counteracted by turbulent diffusion, which tends to stir 
dust particles above the midplane. 
The vertical diffusion coefficient may be parameterized as:
\begin{equation}
D_z=\delta_z\frac{c_s^2}{\Omega},
\end{equation}
where $\delta_z$ is a dimensionless parameter of order the standard $\alpha$ parameter \citep[e.g.][]{Johansenetal2006}. 

  With these notations, equilibrium between diffusion and settling is attained on a timescale
\begin{equation}
t_{\rm vert}=\frac{1}{\Omega\delta_z \mathrm{max}(1,S_z)}=50\:\mathrm{ka}\left(\frac{10^{-4}}{\delta_z\mathrm{max}(1,S_z)}\right)\left(\frac{R}{10\:\mathrm{AU}}\right)^{3/2}
\end{equation}
 and the equilibrium thickness of the dust layer is $H_p=H/\sqrt{1+S_z}$ \citep{Cuzzietal1996}. We have introduced
\begin{eqnarray}
S_z\equiv\frac{\Omega\tau}{\delta_z}&=&\frac{\pi}{2}\frac{\rho_sa}{\Sigma\delta_z}\\&=&20\left(\frac{\rho_sa}{10^{-2}\:\mathrm{kg/m^2}}\right)\left(\frac{10^{-5}}{\delta_z}\right)\left(\frac{10^2\:\mathrm{kg/m^2}}{\Sigma}\right),\nonumber
\end{eqnarray}
where $\tau$ and $\delta_z$ are evaluated at the midplane ($z=0$). $S_z$ is thus a measure of the settling of dust relative to the gas. Significant settling of dust, with interesting feedback on the MRI, corresponds to $S_z\gg 1$. This implies (1) relatively low surface densities (as expected far from the Sun) (2) relatively big grains, say $10^{1\pm 1}\:\mathrm{\mu m}$ (but likely no larger for dust to retain control of the ionization fraction), presumably as a result of coagulation  or (3) a low turbulence level around the midplane, or a combination thereof. Low values of $\delta_z < 10^{-4}$ are seen around the midplane in numerical simulations of layered accretion \citep[e.g.][]{FlemingStone2003,IlgnerNelson2008,OishiMcLow2009,Turneretal2010, OkuzumiHirose2011,Flaigetal2011}
. Note that while \citet{Turneretal2010} did account self-consistently for the dynamics of dust as well as its role on the ionization fraction, no significant effect of dust motion was found (see e.g. their figure 14), but this is because maximum (midplane) values of $S_z$ were only 0.2, 0.8 and 1 for the runs with 1 $\mu$m, 10$\mu$m and 100 $\mu$m grains, respectively\footnote{We identify $\delta_z$ with $\alpha$ evaluated from their figure 7.}.

  With these numerical estimates setting the scales, 
we turn to a reduced model of interaction between 
dust and MRI turbulence.

\section{Reduced model}
\subsection{General Equations}

Consider a vertical section of a protostellar disk. We denote by 
 $y$ a turbulent fluctuation of the gas (with $\epsilon$ still denoting the dust-to-gas ratio). 
Our reduced model consists of the following system of equations:
\begin{equation}
\rho\frac{\partial\epsilon}{\partial t}=\frac{\partial}{\partial z}\left[
\rho\left(Qy^2\frac{\partial\epsilon}{\partial z}-v(z)\epsilon\right)\right]
\label{dn/dt}
\end{equation}
\begin{equation}
\frac{\partial y}{\partial t}=\gamma(\epsilon)y-Ay^3
\label{dy/dt}
\end{equation}
Equation (\ref{dy/dt}) may also be rewritten as:
\begin{equation}
\frac{\partial y^2}{\partial t}=2\gamma(\epsilon)y^2-2Ay^4
\label{dy2/dt}
\end{equation}
The first equation (\ref{dn/dt}) is a diffusion equation for the grains with
a drift term proportional to the velocity $v(z)$.  It has a standard
Fokker-Planck form. The second equation (\ref{dy/dt}) is a Landau equation
\citep{LandauLifshitz1959} for nonlinearly damped fluctuations.  
The constant $Q$ relates $y$ to the
vertical diffusion coefficient $D_z=Qy^2$ and the constant $A$ characterizes
the nonlinear saturation.  Linear growth and damping (here MRI-driven) are embodied in $\gamma$ the 
nominal rate coefficient, which depends on $\epsilon$ by assumptions (i) and (ii) in Section 2.  
The form of
$\gamma(\epsilon)$ is at this point unprescribed, but it is assumed to be
a positive, differentiable quantity.        
Physically, we would expect it to be a monotonically decreasing
function of $\epsilon$. It would reach an asymptotic value for $\epsilon$
below the threshold for recombination on dust grains to be important
(see equation (\ref{gas recombination})), which corresponds to the
dust-free (not necessarily ideal) MHD turbulence value. For $\epsilon$
above another threshold, MRI-powered turbulence is      
suppressed, but other instabilities \citep[e.g.][]{Weidenschilling1980,YoudinGoodman2005,Latteretal2010,LesurPapaloizou2010} 
 may help maintain a minimum
level of turbulence and establish a dynamical equilibrium.
\footnote{The Landau equation used here ignores any transport within the gas, e.g. from the active layers to the dead zone \citep{FlemingStone2003,TurnerSano2008} but were we to add a term $\frac{\partial}{\partial z}\left(Qy^2\frac{\partial y^2}{\partial z}\right)$ in the Landau equation (\ref{dy2/dt}), its ratio with e.g. $Ay^4$ would be of order $Qk^2/A\sim (\Omega/\gamma)(S_z+1)\delta_z\ll 1$ with $k\sim H_p^{-1}$ the reciprocal lengthscale of variation and it would thus be negligible.}


\subsection{Equilibrium solution}
Equilibrium implies:
\begin{equation}
\frac{\partial\mathrm{ln}\epsilon}{\partial z}=\frac{v(z)}{Qy^2}
\label{bckg n}
\end{equation}
\begin{equation}
y^2=\frac{\gamma(\epsilon)}{A}
\label{bckg y}
\end{equation}
Plugging equation (\ref{bckg y}) in equation (\ref{bckg n}) and integrating with respect to $z$ yields:

\begin{equation}
\int_{\epsilon(0)}^\epsilon\frac{\gamma(\epsilon)}{\gamma_{df}}\frac{\mathrm{d}\epsilon}{\epsilon}=-S_{zdf}\left(e^{z^2/2H^2}-1\right),
\label{integrated ODE}
\end{equation}
where we have introduced $\gamma_{df}\equiv\gamma(\epsilon=0)$, the dust-free value of the growth rate, and $S_{zdf}=A\tau(0)c_s^2/\gamma_{df}$ the corresponding value of $S_z$. The value of $\epsilon(0)$ must satisfy:
\begin{equation}
\overline{\epsilon}\equiv\frac{1}{\Sigma}\int_{-\infty}^{+\infty}\rho_p(z)\mathrm{d}z=\frac{1}{\sqrt{2\pi}H}\int_{-\infty}^{+\infty}\epsilon(z)e^{-z^2/2H^2}\mathrm{d}z,
\label{epsilon avg}
\end{equation} 
where we have introduced the dust-to-gas column density ratio $\overline{\epsilon}$. If we approximate $\epsilon(0)\approx\overline{\epsilon}\sqrt{1+S_z}$ (see Section 2.3), where $S_z$ is evaluated at the midplane, this may be replaced by the following simplified relation:
\begin{equation}
S_z=S_{zdf}\frac{\gamma_{df}}{\gamma(\overline{\epsilon}\sqrt{1+S_z})}.
\label{Sz=Sz}
\end{equation}
In principle, depending on the mathematical expression of $\gamma (\epsilon)$ and on the values of $S_{zdf}$ and $\overline{\epsilon}$, it is conceivable that more than one solution exists in terms of $S_z$, or, equivalently, $\epsilon(0)$, or even none if $\gamma$ vanishes too rapidly with increasing $\epsilon$ (which however seems unrealistic, see Section 3.1). 

  We show in Appendix \ref{stability} that \textit{regardless} of the form of $\gamma(\epsilon)$, the equilibrium 
 is 
 linearly \textit{stable}. From equation (\ref{variational principle}) in the appendix, one estimate the damping timescale to be $t_{\rm vert}(\lambda/H_p)^2$ with $\lambda$ the vertical lengthscale of variation of the perturbation: this is basically a diffusion timescale. Physically, if we schematically distinguish between a ``midplane zone'' and an ``atmosphere zone'', we may interpret the lack of instability due to a dust-controlled diffusivity as follows: if the ``atmosphere'' has excess dust, diminishing turbulence there will make the dust flow toward the midplane to cancel the corresponding dust depletion. If, on the other hand, the atmosphere has a dust depletion, enhanced turbulence will soak the excess dust from the midplane through the ``interface'' between the two.

\section{Example}

\begin{figure}
\resizebox{\hsize}{!}{
\includegraphics{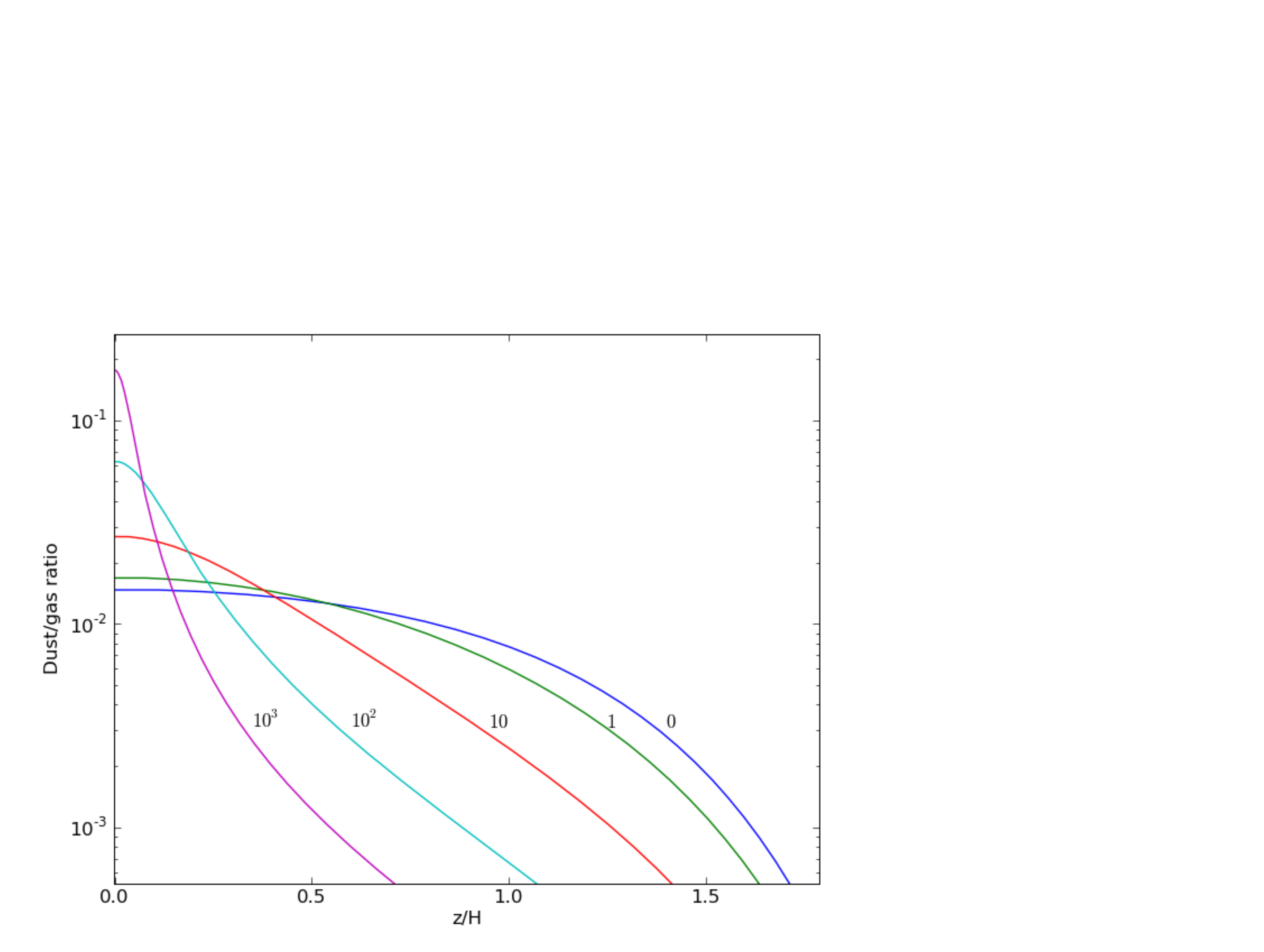}
}
\caption{Plot of the equilibrium dust-to-gas ratio $\epsilon$ profile, assuming the dependence of the growth rate on $\epsilon$ in equation (\ref{Haddock}). Curves are drawn for $S_{zdf}=1$ (i.e., marginal settling for dust-free turbulence levels) for different values of $\epsilon(0)/\epsilon_\ast$ as marked on the figure, assuming a dust/gas column density ratio of $10^{-2}$. 
}
\label{background}
\end{figure}

\begin{figure}
\resizebox{\hsize}{!}{
\includegraphics{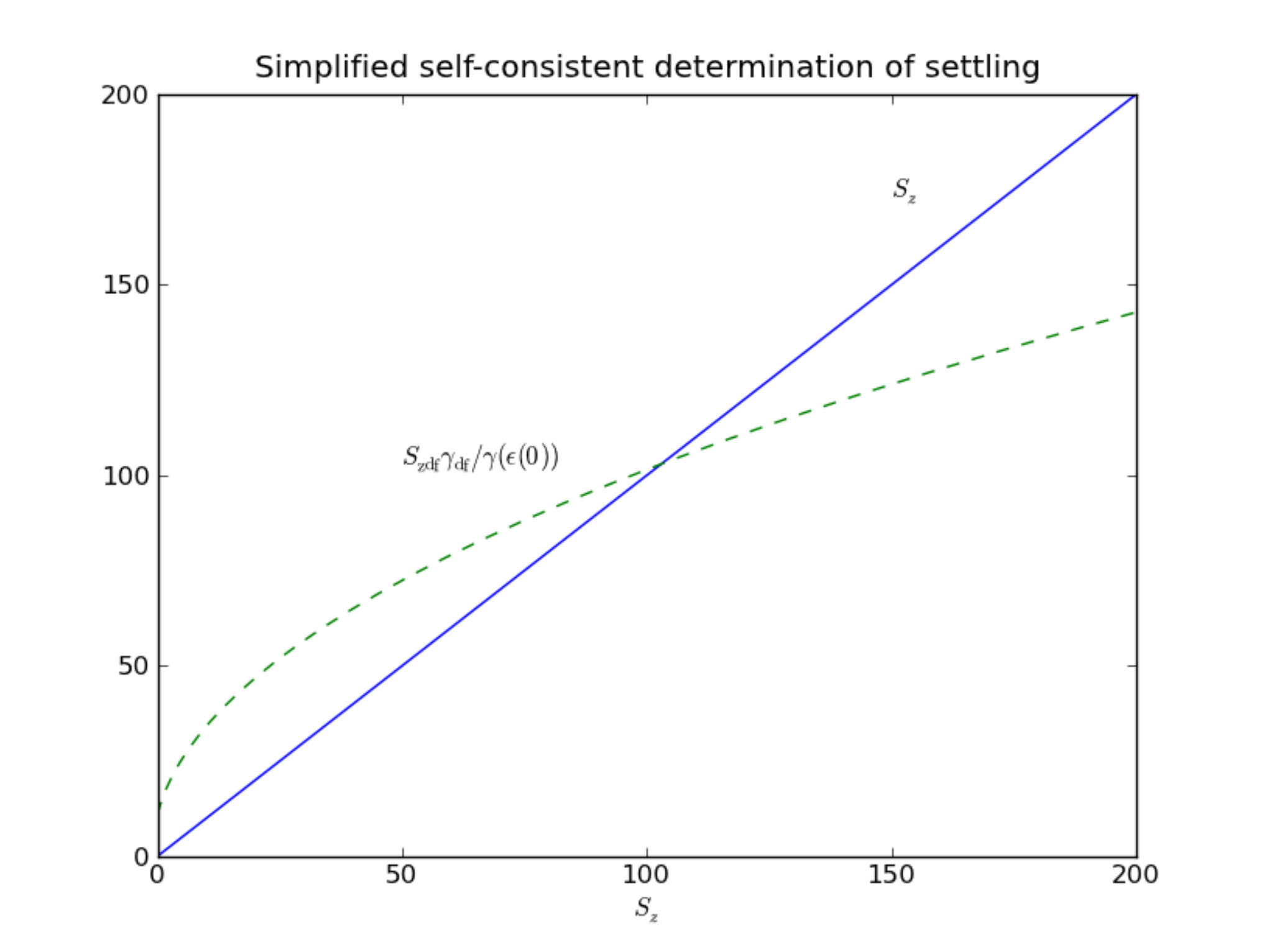}
}
\caption{Plot of the left-hand-side (continuous) and the right-hand-side (dashed) of the simplified equation (\ref{Sz=Sz}) for the example functional dependence of the growth rate displayed in equation (\ref{Haddock}). For any parameter, there is always one unique solution (corresponding to the intersection between the two curves) for this functional dependence. Here, $\overline{\epsilon}/\epsilon_\ast=10$ and $S_{zdf}=1$.}
\label{Sz self-consistent}
\end{figure}

\begin{figure}
\resizebox{\hsize}{!}{
\includegraphics{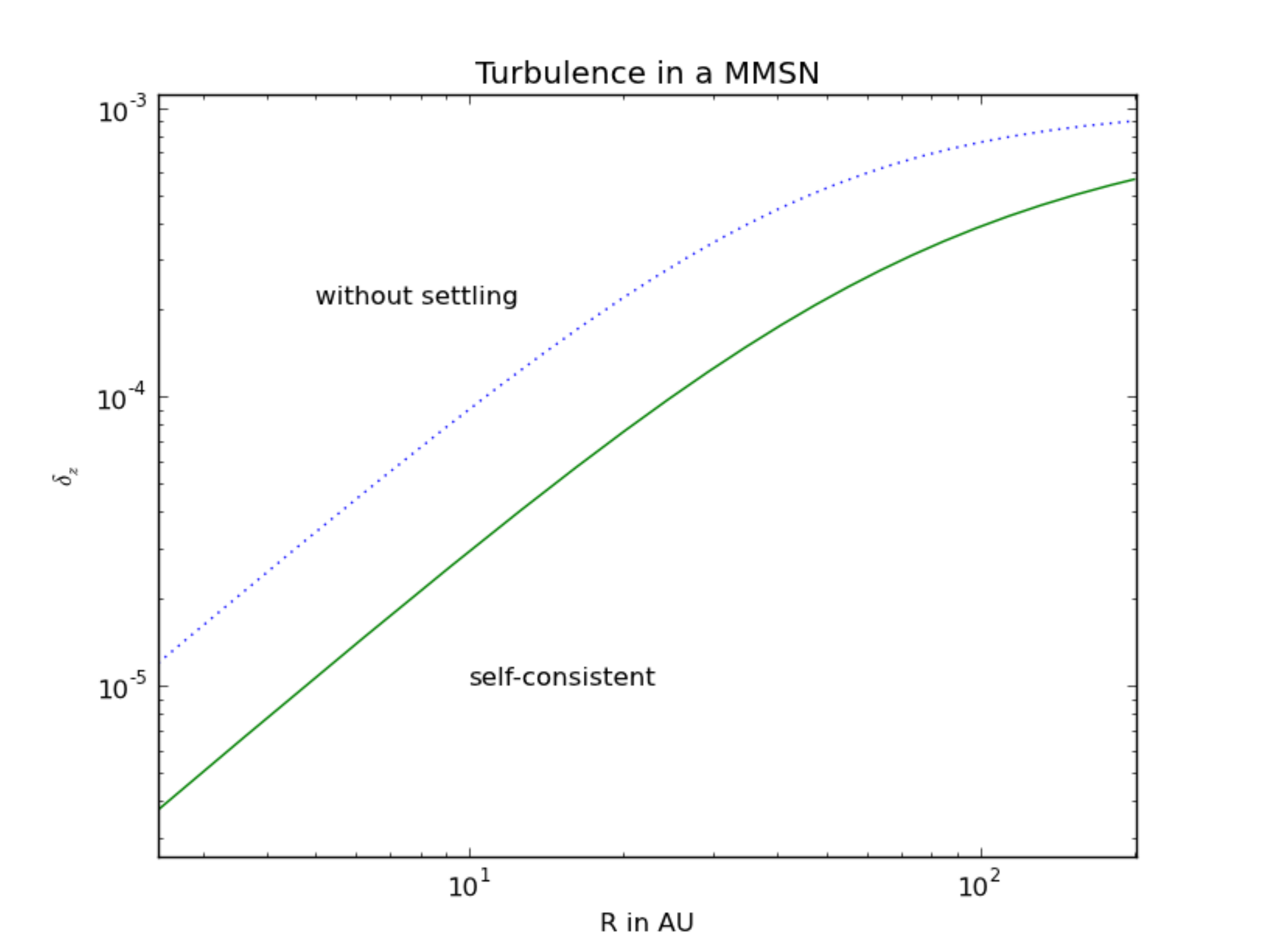}
}
\caption{Radial variation of the turbulence parameter $\delta_z$ (evaluated at the midplane) in a MMSN assuming that the growth rate is given by equation (\ref{Haddock}). We assume a constant dust-to-gas column density ratio $\overline{\epsilon}=0.01$, $\rho_sa=0.1\:\mathrm{kg/m^2}$ (i.e. 30 $\mu$m-radius grains), $\delta_{zdf}=10^{-3}$ and take $\epsilon_\ast=10^{-3}(R/10\:\mathrm{AU})^{3/2}$ (to mimick the dependence on $\Omega$ of dimensionless numbers pertaining to nonideal MHD in Section 2.2). The dotted line is the estimate ignoring vertical settling and the continuous line is the self-consistent estimate using simplified equation (\ref{Sz=Sz}).}
\label{deltaz MMSN}
\end{figure}

As an example, consider 
 the following 
functional dependence for the growth rate:
\begin{equation}
\gamma(\epsilon)=\frac{\gamma_{df}}{1+\epsilon/\epsilon_\ast}.
\label{Haddock}
\end{equation}
One may think of $\epsilon_\ast$ as the critical value of $\epsilon$ for which the relevant dimensionless number of Section 2.2 is at its threshold value. The asymptotic value of the growth rate for $\epsilon\gg\epsilon_\ast$ is zero, i.e. we ignore any ``background'' hydrodynamical turbulence. 

  We obtain from equation (\ref{integrated ODE}):
\begin{equation}
\epsilon(z)=\left(\left(\frac{1}{\epsilon_\ast}+\frac{1}{\epsilon(0)}\right)\exp{\left(S_{zdf}(e^{z^2/2H^2}-1)\right)}-\frac{1}{\epsilon_\ast}\right)^{-1}.
\end{equation}
This is plotted in Fig. \ref{background}. It is seen that enhanced dust abundance at the midplane entails an enhanced settling efficiency, hence a peaked distribution. But as $|z|$ increases and thus the dust fraction drops, turbulence can increase and the profile becomes shallower, until dust no longer affects the turbulence level, by which altitude one has:
\begin{equation}
\epsilon(z)\approx \frac{\exp{(-S_{zdf}(e^{z^2/2H^2}-1))}}{\frac{1}{\epsilon_\ast}+\frac{1}{\epsilon(0)}},
\end{equation} 
that is, $(1+\epsilon(0)/\epsilon_\ast)^{-1}$ times the result for a vertically constant $\delta_z$ \citep[e.g.][]{FromangNelson2009}, a depletion due to the aforementioned enhanced dust concentration at the midplane. If $\epsilon (0) \gg \epsilon_\ast$, the self-consistently determined dust concentration profile sets the vertical extent of the dead zone and the active layers.

  The mass balance constraint on dust expressed by equation (\ref{epsilon avg}) reads here:
\begin{equation}
\overline{\epsilon}=\frac{2}{\sqrt{\pi}}\int_0^{+\infty}\frac{e^{-x^2}}{\left(\frac{1}{\epsilon_\ast}+\frac{1}{\epsilon (0)}\right)\mathrm{exp}\left(S_{zdf}(e^{x^2}-1)\right)-\frac{1}{\epsilon_\ast}}\mathrm{d}x
\end{equation}
Since the right-hand-side is a monotonic function of $\epsilon (0)$, increasing from 0 to $+\infty$, there is always one unique solution, given $S_{zdf}$ and $\overline{\epsilon}$.

  The simplified equation (\ref{Sz=Sz}), illustrated graphically in Fig. \ref{Sz self-consistent},  may be solved for $S_z$ as:
\begin{eqnarray}
S_z=S_{zdf}\Bigg[1+\frac{S_{zdf}}{2}\left(\frac{\overline{\epsilon}}{\epsilon_\ast}\right)^2\nonumber\\+\sqrt{(1+S_{zdf})\left(\frac{\overline{\epsilon}}{\epsilon_\ast}\right)^2+\frac{S_{zdf}^2}{4}\left(\frac{\overline{\epsilon}}{\epsilon_\ast}\right)^4}\Bigg]
\end{eqnarray}


If $\overline{\epsilon} \ll \epsilon_\ast$ and $S_{zdf}(\overline{\epsilon}/\epsilon_\ast)^2\ll 1$, there is essentially no effect of dust and $S_z\approx S_{zdf}$. At the other extreme, if  $S_{zdf}\mathrm{max}\left((\overline{\epsilon}/\epsilon_\ast),(\overline{\epsilon}/\epsilon_\ast)^2\right) \gg 1$, we have:
\begin{equation}
S_{zdf}\approx \left(S_{zdf}\frac{\overline{\epsilon}}{\epsilon_\ast}\right)^2=\left(\frac{\pi}{2}\frac{\overline{\epsilon}}{\Sigma\delta_{zdf}c_\ast}\right)^2,
\end{equation}
where we have set $\epsilon_\ast=c_\ast\rho_sa$, with $c_\ast$ independent of dust properties, in order to account for the dependence of the ionization fraction on grain size (see Section 2.1). We see that, in this limit, the dependences of settling on grain size cancel out. This is because for a given dust concentration, larger grains drift more rapidly toward the midplane, but at the same time allow higher turbulence levels. Certainly, this (asymptotically) exact cancellation is specific to the dependence we have chosen, but the simulations of \citet{Turneretal2010} show only a weak dependence of $S_z$ on size, with the former varying by a factor of $\sim$5 despite a two-order-of-magnitude variation of the latter (see Section 2.3). 

  Extrapolating for a range of heliocentric distances, we have plotted the radial profile of $\delta_z$ \textit{evaluated at the midplane} in figure \ref{deltaz MMSN}. It is seen that ignoring settling of grains leads to an overestimate of $\delta_z$ and hence an underestimate of the heliocentric distance of the 
 outer edge of the dead zone. This is because settling induces larger dust concentrations at the miplane, compared to the perfect vertical mixing assumption, and thus lower turbulence levels there. As to the \textit{vertically averaged} $\delta_z$---a proxy for the standard $\alpha$ parameter---ignoring settling would lead to an underestimate of its value, since $\gamma$ is here a convex function of $\epsilon$, but in general, the effect of settling on this average depends on the particular mathematical form of $\gamma(\epsilon)$.

\section{Conclusion}
  We have studied the interaction between MRI turbulence and dust grains, 
allowing the dust to control the ionization level (and thus MHD turbulence 
level), while turbulent fluctuations inhibit dust grain settling to 
the disk midplane.  We have used a reduced model consisting of two 
coupled equations: a Landau equation for the turbulent fluctuation 
amplitude and a Fokker-Planck equation governing the vertical dynamics of 
the grains via a fluctuation-dependent diffusion coefficient.  
Unconditionally stable equilibrium solutions for the vertical grain 
distribution were found. Compared with models in which the turbulent 
fluctuation level is constant with height, the solutions were signficantly 
more concentrated near the midplane, with a lower level of MHD turbulence 
in this region.

  From simple estimates of the ionization fraction and the settling
parameter, we found that the interaction studied here is most relevant
near the outer edge of the disk's dead zone.  Indeed, the grains are
likely to determine the dead zone morphology, and will likely
control the region's heliocentric extent. Also, an enhanced settling could lead to flatter disks and hence steeper decreases of temperature with heliocentric distance.  Thus, the model may
be used in conjunction with numerical simulations (global or local),
to sharpen observational predictions that will be of great interest
when ALMA becomes fully operational.   

\section*{Acknowledgments}
We thank the anonymous referee for his/her review and for bringing some additional caveats to our model.

\bibliographystyle{aa}
\bibliography{bibliography}

\begin{appendix}

\section{Estimation of the molecule/metal ion ratio}
\label{molecule metal}
In this appendix, we provide some justification of the normalization chosen for $\beta_{\rm eff}$ in Section 2.1. To that end, we estimate the ratio between the number density $n_{m^+}$ of molecular ions ($m^+$, e.g. HCO$^+$) and that $n_{M^+}$ of metal ions ($M^+$, e.g. Mg$^+$, Na$^+$). We thus need to refine the model in the main text by distinguishing between the two. The resulting governing equations are the same as \citet{Fromangetal2002}, whose notation we adopt, with the addition of grains:
\begin{equation}
\frac{\partial n_{m^+}}{\partial t}=\zeta n_n - \beta n_en_{m^+} -\beta_tn_Mn_{m^+}- I_{m^+}\pi a^2 v_{Tm^+}n_pn_{m^+}
\label{dnm+/dt}
\end{equation}
\begin{equation}
\frac{\partial n_{M^+}}{\partial t}=\beta_tn_Mn_{m^+}-\beta_rn_{M^+}n_e- I_{M^+}\pi a^2 v_{TM^+}n_pn_{M^+},
\label{dnM+/dt}
\end{equation}
with $n_M$ the metal number density, $\beta = 3\times 10^{-13}$ $(100\:\mathrm{K}/T)^{1/2} \:\mathrm{m^3/s}$ the dissociative recombination rate coefficient for molecular ions, $\beta_r=3\times 10^{-18}\:(100\:\mathrm{K}/T)^{1/2} \:\mathrm{m^3/s}$ the radiative recombination rate coefficient for metal atoms and $\beta_t=3\times 10^{-15}\:\mathrm{m^3/s}$ the rate coefficient of charge transfer from molecular ions to metal atoms.

 Equation (\ref{dni/dt}) is retrieved by summing these equations, recalling that $n_i=n_{m^+}+n_{M^+}$, 
 if one puts:
\begin{equation}
\frac{1}{\sqrt{m_i}}=\frac{1}{n_i}\left(n_{M^+}\frac{1}{\sqrt{m_{M^+}}}+n_{m^+}\frac{1}{\sqrt{m_{m^+}}}\right)
\end{equation}
\begin{equation}
I_{i}=\sqrt{m_i}\left(\frac{I_{M^+}}{\sqrt{m_{M^+}}}\frac{n_{M^+}}{n_i}+\frac{I_{m^+}}{\sqrt{m_{m^+}}}\frac{n_{m^+}}{n_i}\right)
\end{equation}
\begin{equation}
\beta_{\rm eff} = \frac{n_{M^+}}{n_i}\beta_r+\frac{n_{m^+}}{n_i}\beta
\end{equation}
At equilibrium, provided that
\begin{equation}
\frac{\beta_rn_M^+n_e}{I_{M^+}v_{TM^+}\pi a^2 n_pn_M^+}=\frac{4}{3}\frac{x_e\beta_r\rho_sa}{\epsilon I_{M^+}v_{TM^+}m_{H_2}}\ll 1,
\end{equation}
that is, if we plug in equation (\ref{xe}) for $x_e$,
\begin{equation}
\left(\frac{\rho_sa}{\epsilon}\right)^2\frac{\sqrt{m_em_{M^+}}}{I_eI_{M^+}m_{H_2}^2}\frac{\zeta\beta_r}{P}\ll 1,
\end{equation}
which, given that $\beta_r\ll\beta$, is essentially guaranteed by inequality (\ref{gas recombination}), we draw from equation (\ref{dnM+/dt}):
\begin{eqnarray}
\frac{n_{m^+}}{n_{M^+}}=\frac{3}{4}\frac{I_{M^+}v_{TM^+}m_{H_2}\epsilon}{\rho_sa\beta_tx_M}=10^{-3}I_{M^+}\left(\frac{T}{100\:\mathrm{K}}\right)^{1/2}\left(\frac{\epsilon}{10^{-2}}\right)\nonumber\\\left(\frac{10^{-2}\:\mathrm{kg/m^2}}{\rho_sa}\right)\left(\frac{10^{-7}}{x_M}\right)
\end{eqnarray}
where $x_M\equiv n_M/n_n$. It is normalized to a reasonable value considering the solar abundances for e.g. Na, Mg, Si, K, Fe \citep[$\mathrm{log}(M/H)= -5.70, -4.45, -4.46, -6.89, -4.53$, respectively ;][]{Lodders2003}  and the expected 1-2 order-of-magnitude depletion due to condensation (see Section 2.1). Then, to a good approximation (if that ratio is larger than 10$^{-5}$), $\beta_{\rm eff}\approx (n_{m^+}/n_{M^+})\beta$, justifying the normalization chosen for $\beta_{\rm eff}$ in inequality (\ref{gas recombination}). The latter then becomes:
\begin{eqnarray}
 \frac{\beta_{\rm eff} n_in_e}{\zeta n_n}&\approx & \frac{\sqrt{m_em_i}}{m_{H_2}}\frac{I_{M^+}}{I_eI_i}\frac{\beta}{\beta_t}\frac{v_{TM^+}}{x_M}\frac{\rho_sa}{\epsilon}\frac{\zeta}{P}\nonumber\\
& \approx & 10^{-4}\left(\frac{\zeta}{10^{-17}\:\mathrm{s^{-1}}}\right)\left(\frac{10^{-3}\:\mathrm{Pa}}{P}\right)\left(\frac{10^{-2}}{\epsilon}\right)\nonumber\\&&\left(\frac{\rho_sa}{10^{-2}\:\mathrm{kg/m^2}}\right)\left(\frac{10^{-7}}{x_M}\right)\nonumber\\
& \ll & 1
\end{eqnarray}

\section{Linear stability analysis}
\label{stability}
We consider the behavior of the equilibrium solution
of our reduced model system
(\ref{dn/dt}) and (\ref{dy2/dt})
to linear perturbations 
with a time dependence of the form $\exp{(st)}$. 
The linearized form of equations (\ref{dn/dt}) and (\ref{dy2/dt}) is:
\begin{equation}
s\rho\delta\epsilon = \frac{\partial}{\partial z}\left(\rho\left(Qy^2\frac{\partial\delta\epsilon}{\partial z}+Q\frac{\partial\epsilon}{\partial z}\delta y^2-v(z)\delta\epsilon\right)\right)
\label{dn/dt perturbed}
\end{equation}
\begin{equation}
s\delta y^2=2\gamma'(\epsilon)y^2\delta\epsilon +2\left(\gamma(\epsilon)- 2Ay^2\right)\delta y^2.
\label{dy2/dt perturbed}
\end{equation}
Using background equation (\ref{bckg y}), 
equation (\ref{dy2/dt perturbed}) yields:
\begin{equation}
\delta y^2=\frac{\gamma'(\epsilon)y^2\delta\epsilon}{\gamma(\epsilon)+s/2},
\label{deltay2}
\end{equation}
Then, using equation (\ref{bckg n}), and provided 
  $2\gamma+s\neq 0$, equation (\ref{dn/dt perturbed}) may be rewritten as: 
\begin{equation}
s\rho\delta\epsilon = \frac{\partial}{\partial z}\left(\rho\frac{Qy^2}{h}\frac{\partial}{\partial z}\left(h\delta\epsilon\right)\right)
\label{dn/dt perturbed with h}
\end{equation}
With\footnote{We use the fact that
\begin{eqnarray}
\frac{d}{dt}
\left(\mathrm{ln}|t-w|+i\mathrm{arctan}\frac{t-\mathrm{Re}w}{\mathrm{Im}w}\right)=\frac{1}{t-w}\nonumber,
\end{eqnarray}
for $t$ a real variable and $w$ a complex constant.}:
\begin{equation}
h=
\left\{\begin{array}{rr}
\frac{1}{\epsilon}|\gamma+\frac{s}{2}|\:\mathrm{if}\:\mathrm{Im}s= 0\\
\frac{1}{\epsilon}|\gamma+\frac{s}{2}| \exp{\left(-i\mathrm{arctan}\frac{2\gamma+\mathrm{Re}s}{\mathrm{Im}s}\right)} \:\mathrm{if}\:\mathrm{Im}s\neq 0
\end{array}\right.
\label{h}
\end{equation}
where Re and Im denote the real and imaginary parts, respectively. If we mutliply equation (\ref{dn/dt perturbed with h}) by $(h\delta\epsilon)^\ast$ and integrate over $z$, we obtain:
\begin{equation}
s=-\frac{\int_{-\infty}^{+\infty}\frac{Qy^2}{h}|\frac{\partial}{\partial z}\left(h\delta\epsilon\right)|^2\rho\mathrm{d}z}{\int_{-\infty}^{+\infty}h^\ast|\delta\epsilon|^2\rho\mathrm{d}z}
\label{variational principle}
\end{equation}
We now want to show that this equation implies that the real part of $s$ (i.e. the growth rate of the perturbation) is negative, i.e. that the system is linearly stable. We distinguish two cases:

\textit{Case 1: $\mathrm{Im}s=0$}. In this case $h$ is real and positive. It is then clear from equation (\ref{variational principle}) that $s$ must be real and negative\footnote{Recall that in writing equation (\ref{variational principle}), we had assumed that $2\gamma+s$ was nonzero everywhere; if that is not the case, it is clear that $s$ would be a negative real number in that case too.}. 

\textit{Case 2: $\mathrm{Im}s\neq 0$}. We proceed \textit{ab absurdo}, by supposing $\mathrm{Re}s > 0$. Then, 
$2\gamma+\mathrm{Re}s$ is always strictly positive, from which it follows, from the definition of $h$ (equation (\ref{h})) that:
\begin{equation}
\left\{\begin{array}{rr}
-\frac{\pi}{2}<\mathrm{arg}h<0\:\mathrm{if}\:\mathrm{Im}s>0\\
0<\mathrm{arg}h<\frac{\pi}{2}\:\mathrm{if}\:\mathrm{Im}s<0
\end{array}\right.
\end{equation}
where $\mathrm{arg}h$ is the real number (between $-\pi$ and $\pi$) satisfying $h=|h|\mathrm{exp}(i\mathrm{arg}h)$. In the complex plane, for a given eigenvector (and thus a fixed $\mathrm{Im}s$), $h$ is thus confined to a determined quadrant.  
Thus, the integrand in each of the integrals of (\ref{variational principle}) is always in the symmetric quadrant with respect to the real axis (because $\mathrm{arg}(1/h)=\mathrm{arg}(h^\ast)=-\mathrm{arg}h$). Same holds for the integrals themselves. It follows that the \textit{phase} of the \textit{ratio} of these integrals, which equals the \textit{difference} of the \textit{phases} of these two integrals, cannot be larger than $\pi/2$ in absolute value. In other words, the ratio of the integrals has a positive real part and hence $\mathrm{Re}s<0$. 
This contradicts the hypothesis and $s$ must thus have a negative real part.


\end{appendix}

\end{document}